\documentclass{aa}  

\usepackage{graphicx}
\usepackage{txfonts}
\usepackage[colorlinks,citecolor=blue]{hyperref}
\begin{document} 

\title{Formation of giant plasmoids at the pulsar wind termination shock: A possible origin of the inner-ring knots in the Crab Nebula}

\titlerunning{Formation of giant plasmoids at the pulsar wind termination shock}

\author{Beno\^it Cerutti \inst{1}\and Gwenael Giacinti\inst{2}}

\institute{Univ. Grenoble Alpes, CNRS, IPAG, 38000 Grenoble, France\\
           \email{benoit.cerutti@univ-grenoble-alpes.fr}
           \and
           Max-Planck-Institut f\"ur Kernphysik, Postfach 103980, 69029 Heidelberg, Germany\\
           \email{gwenael.giacinti@mpi-hd.mpg.de}
           }

\date{Received \today; accepted \today}

 
\abstract
{Nearby pulsar wind nebulae exhibit complex morphological features: jets, torus, arcs, and knots. These structures are well captured and understood in the scope of global magnetohydrodynamic models. However, the origin of knots in the inner radius of the Crab Nebula remains elusive.}
{In this work, we investigate the dynamics of the shock front and downstream flow with a special emphasis on the reconnecting equatorial current sheet. We examine whether giant plasmoids produced in the reconnection process could be good candidates for the knots.}
{To this end, we perform large semi-global three-dimensional particle-in-cell simulations in a spherical geometry. The hierarchical merging plasmoid model is used to extrapolate numerical results to pulsar wind nebula scales.}
{The shocked material collapses into the midplane, forming and feeding a large-scale, but thin, ring-like current layer. The sheet breaks up into a dynamical chain of merging plasmoids, reminiscent of three-dimensional reconnection. Plasmoids grow to a macroscopic size. The final number of plasmoids predicted is solely governed by the inverse of the dimensionless reconnection rate.}
{The formation of giant plasmoids is a robust feature of pulsar wind termination shocks. They provide a natural explanation for the inner-ring knots in the Crab Nebula, provided that the nebula is highly magnetized.}

               
\maketitle


\section{Introduction}

High-resolution X-ray images of nearby pulsar wind nebulae, as best illustrated by the iconic \emph{Chandra} images of the Crab Nebula \citep{2000ApJ...536L..81W, 2002ApJ...577L..49H, 2004ApJ...609..186M} and of the Vela pulsar wind nebula \citep{2001ApJ...556..380H, 2001ApJ...552L.129P, 2013ApJ...763...72D}, have uncovered complex morphological features. This played a crucial role in the recent development of pulsar wind nebula theory (see, e.g., \citealt{2015SSRv..191..391K} for a review and references therein) and, more generally, the study of relativistic magnetized outflows.

\begin{figure}
\centering
\includegraphics[width=\hsize]{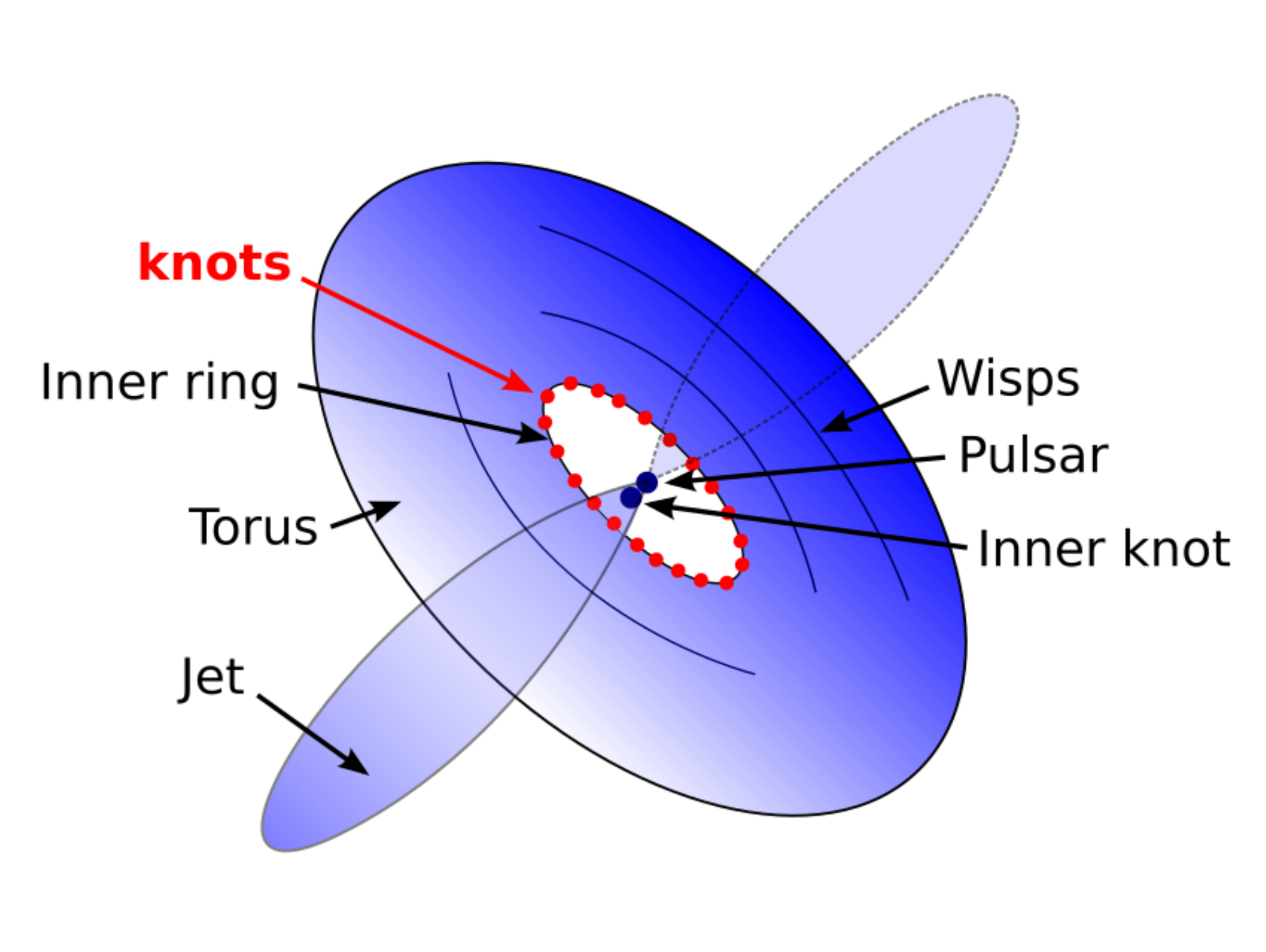}
\caption{Diagram schematically representing the morphological features observed in the Crab Nebula in the optical and X-ray bands. The MHD model captures all these features well, except the knots along the inner ring (in red), whose origin is unknown. Solving this mystery is the main focus of this work.}
\label{fig0}
\end{figure}

At the largest scales, a pulsar wind nebula can be well described as being composed of a jet-like structure aligned along the pulsar rotation axis and of a bright torus-like structure lying in the equatorial plane. The pulsar is seemingly located at the center of the torus and near the base of the jet. At smaller scales, and currently only visible in the Crab and Vela pulsar wind nebulae, one can distinguish bright thin, arc-like structures within the torus that are moving away from the pulsar, also known as the ``wisps'' \citep{1969ApJ...156..401S}, and an ``inner ring'' located at the inner edge of the torus \citep{2000ApJ...536L..81W}. A closer look at the Crab inner ring reveals that it is in fact composed of a series of about two dozen bright compact knots (\citealt{2000ApJ...536L..81W, 2002ApJ...577L..49H}), and at least two knots are also reported in Vela (\citealt{2001ApJ...554L.189P}). These knots are highly variable, although they seem stationary and confined within the inner ring. They appear as unpolarized structures in the optical band \citep{2008ARA&A..46..127H}, and they should not be confused with the ``inner knot.'' The latter is another salient feature of the Crab Nebula discovered with the \emph{Hubble Space Telescope} in the optical \citep{1995ApJ...448..240H}; it is highly variable and polarized \citep{2013MNRAS.433.2564M}, but it is too close to the pulsar projected on the sky to be singled out as a separate component in X-rays. Figure~\ref{fig0} schematically represents all of the aforementioned morphological features.

It is now well understood that the jet-torus structure stems from the interaction of the anisotropic relativistic magnetized wind blown by the pulsar with the surrounding medium \citep{2002MNRAS.336L..53B, 2002MNRAS.329L..34L}. A shock forms from this interaction, leading to compression and heating of the pulsar wind and (re)acceleration of the relativistic electron-positron pairs it contains; this gives rise to bright synchrotron radiation \citep{1974MNRAS.167....1R, 1984ApJ...283..710K, 1984ApJ...283..694K}. Two-dimensional (2D) axisymmetric magnetohydrodynamic (MHD) simulations have successfully reproduced the observed jet-torus structure, which was identified as the plasma downstream of the pulsar wind termination shock \citep{2003MNRAS.344L..93K, 2004Ap&SS.293..107K, 2004A&A...421.1063D}. In this framework, the moving wisps result from the feedback of the highly dynamical downstream flow on the shock front location \citep{2009MNRAS.400.1241C}, while the inner knot originates from the Doppler-boosted emission of the termination shock pointing toward the observer \citep{2004Ap&SS.293..107K, 2011MNRAS.414.2017K}. Three-dimensional (3D) MHD simulations confirmed and accurately reproduced the morphology and time evolution of the Crab Nebula down to the smallest details \citep{2013MNRAS.431L..48P, 2014MNRAS.438..278P}, giving us high confidence in this physical model.

The morphological features that are not captured in the scope of the 3D MHD model are the brightness of the inner ring and its knots \citep{2014MNRAS.438..278P}. This region is often regarded as the location of the termination shock itself in the equatorial plane \citep{2000ApJ...536L..81W}. Yet, it stands out from the rest of the nebula because this is where both polarities of the (mostly toroidal) magnetic field meet and reconnect \citep{2018ApJ...863...18G}. Using 2D particle-in-cell (PIC) simulations, we have shown in a previous study that the anisotropy of the wind naturally results in the formation of a large-scale, but thin, reconnecting current sheet in the midplane, which plays a key role in particle acceleration in the nebula (as well as shear flow acceleration; see \citealt{2020A&A...642A.123C}). Therefore, the equatorial current sheet is a good candidate for explaining the brightness of the inner ring, but its knotty nature could not be addressed with these 2D simulations. 

It is now well established that a long, thin current sheet is prone to the tearing and plasmoid instabilities \citep{1979SvA....23..460Z, 2007PhPl...14j0703L, 2010PhRvL.105w5002U} that lead to the fragmentation of the sheet into multiple secondary current layers separated by plasma over-densities trapped in magnetic islands, or flux ropes in 3D. As reconnection proceeds, magnetic islands merge to form bigger structures, eventually giving rise to giant, macroscopic plasmoids \citep{2012PhPl...19d2303L} filled with energetic particles. In this work, we investigate the dynamics of the shock front and downstream flow in light of more realistic semi-global 3D PIC simulations in a spherical geometry. We draw special attention to the development of flux ropes and examine whether these structures could be good candidates for explaining the observed knots along the inner ring of the Crab Nebula. In Sect.~\ref{sect_setup} we describe our numerical setup, which was inspired from \citet{2020A&A...642A.123C} and adapted to a spherical geometry. In Sect.~\ref{sect_results} we present the dynamics of the shock, with an emphasis on the formation of the equatorial current sheet and the flux ropes it contains. We provide a detailed analysis of their evolution and propose a toy model that can be used to apply and extrapolate our results to any pulsar wind nebula.

\section{Numerical setup}\label{sect_setup}

We worked with the {\tt Zeltron} PIC code using 3D spherical coordinates \citep[$r,\theta,\phi$;][]{2013ApJ...770..147C, 2016MNRAS.457.2401C}. The numerical domain is a spherical wedge of the inner radius, $r_{\rm min}$, and outer radius, $r_{\rm max}=10 r_{\rm min}$, limited by the polar angle ranging from $\theta_{\rm min}=\pi/4$ to $\theta_{\rm max}=3\pi/4$, and azimuth ranging from $\phi_{\rm min}=0$ to $\phi_{\rm max}=\pi/2$. The polar axis ($\theta=0$) is aligned along the pulsar rotation axis. We removed the polar regions to concentrate our computational resources on the equatorial regions. We verified that this choice does not significantly affect the dynamics of the flow in the equatorial belt and the results presented in this work. All runs were composed of ($1600\times1024\times1024$) cells along the $r$, $\theta$, and $\phi$ directions, respectively. The grid cells were equally spaced in $\log r$, $\cos\theta$, and $\phi$ to optimize the numerical resolution. The outer radius was a perfectly reflecting boundary with no loss of energy for both the electromagnetic fields and the particles. The $\theta$ boundaries were periodic for the particles but reflecting for the fields, and the $\phi$ boundaries were periodic for both the fields and the particles.

The numerical box was initialized with a perfectly radial wind composed exclusively of electron-positron pairs propagating outward with a bulk velocity of $V_0/c=0.99$, or a bulk Lorentz factor of $\Gamma_0\approx 7$. This choice satisfies both the need to model an ultra-relativistic flow, as found in pulsar winds (i.e., $\Gamma_0\gg 1$), and the need to limit spurious heating of the beam due to numerical Cherenkov radiation, which limits the maximum Lorentz factor in the simulation to $\Gamma\lesssim 10$. The shock forms when the initial beam of plasma hits and reflects on the radial outer boundary, giving rise to two identical but counter-propagating beams (e.g., \citealt{2013ApJ...771...54S}). This setup is numerically convenient and easy to implement but has the disadvantage of producing a reverse shock that propagates inward; this is not the case in the Crab Nebula, where it is approximately stationary in the frame of the observer. The outer boundary can be seen as the contact discontinuity separating the pulsar wind material from the ambient medium, which is not simulated here. Therefore, the numerical experiment presented here is valid in the frame of the contact discontinuity of the flow, which is presumably still moving outward in young pulsar wind nebulae such as the Crab.

This far from the pulsar magnetospheric regions where the wind is launched, the magnetic field in the upstream medium entering the shock can be considered as purely toroidal \citep{1973ApJ...180L.133M, 1999A&A...349.1017B}. We further assumed that the oscillating component of the field shaped by the pulsar rotation when the rotation and magnetic axes of the star are misaligned at an angle $\chi\neq 0$ (i.e., the striped wind; \citealt{1990ApJ...349..538C, 1994ApJ...431..397M}) has fully dissipated when it reaches the termination shock \citep{1990ApJ...349..538C, 2017A&A...607A.134C, 2020A&A...642A.204C}. A good proxy for the field is given by\footnote{Assuming that $\chi\leq \pi/2$. If $\chi>\pi/2$, then $B_{\phi}\rightarrow -B_{\phi}$ in Eq.~(\ref{eq_bphi}).} \citep{2020A&A...642A.123C}
\begin{equation}
B_{\phi}=B_0\left(\frac{r_{\rm min}}{r}\right)\tanh\left(\frac{\theta-\pi/2}{\chi}\right)\sin\theta,
\label{eq_bphi}
\end{equation}
$B_{r}=0$ and $B_{\theta}=0$. The $\sin\theta/r$ term is borrowed from the split-monopole model \citep{1973ApJ...180L.133M, 1999A&A...349.1017B}, a good description of the asymptotic dissipation-less pulsar wind structure, while the $\tanh((\theta-\pi/2)/\chi)$ term translates the full dissipation of the striped wind component. The electric field in the wind is set by the ideal Ohm's law, $\mathbf{E}=-\mathbf{V}\times\mathbf{B}/c$, giving a single non-vanishing component, $E_{\theta}=V_0 B_{\phi}/c$. 

In the steady state, the wind must carry the current density $\mathbf{J}=c/(4\pi)\boldsymbol{\nabla}\times\mathbf{B}$. Using Eq.~(\ref{eq_bphi}) yields
\begin{equation}
J_{r}=\frac{cB_0}{4\pi}\frac{r_{\rm min}}{r^2}\left\{\frac{\sin\theta}{\chi\cosh^2\left(\frac{\theta-\pi/2}{\chi}\right)}+2\tanh\left(\frac{\theta-\pi/2}{\chi}\right)\cos\theta\right\},
\end{equation}
$J_{\theta}=0$ and $J_{\phi}=0$. Ohm's law combined with Gauss' law indicates that the wind has a net charge, given by the following charge density,
\begin{equation}
\rho=\frac{1}{4\pi}\boldsymbol{\nabla}\cdot\mathbf{E}=\frac{V_0}{c^2}J_{r}=\rho_+ +\rho_-,
\end{equation}
where 
\begin{equation}
\rho_+=\frac{V_0 B_0}{4\pi c}\frac{r_{\rm min}}{r^2}\frac{\sin\theta}{\chi\cosh^2\left(\frac{\theta-\pi/2}{\chi}\right)}\geq 0,
\end{equation}
\begin{equation}
\rho_-=\frac{V_0 B_0}{2\pi c}\frac{r_{\rm min}}{r^2}\tanh\left(\frac{\theta-\pi/2}{\chi}\right)\cos\theta\leq 0.
\end{equation}
Following \citet{2020A&A...642A.123C}, we assumed that the positive charge density, $\rho_+$, is carried away by positrons only of number density $n_+=\rho_+/e$ (where $e$ is the elementary electric charge) and $\rho_-$ by electrons of density $n_-=-\rho_-/e$, both species propagating outward in the wind at the speed $V_0$. The pulsar wind must also fulfill quasi-neutrality (i.e., $|\rho|\ll ne$). This condition is easily achieved by adding a uniform plasma density of $n_{\rm w}=n_0 (r_{\rm min}/r)^2$ to both the electrons and positrons, that is drifting outward at $V_0$ such that the total electron density is $n_{\rm e}=n_{\rm w}+n_-$ and the total positron density is $n_{\rm p}=n_{\rm w}+n_+$. The injected plasma density profile in the wind is modeled by four particles per cell. Particles are continuously added to the wind zone from an injector that is initially located at $r=0.95 r_{\rm max}$ and then recedes at the speed of light toward the inner boundary as the simulation continues. The run stops once the injector has reached the inner boundary, that is, $t_{\rm max}=8.5 r_{\rm min}/c$; this represents about $12,000$ time steps, which are fixed at half the Courant-Friedrich-Lewy stability condition.

The fiducial plasma density, $n_0$, and magnetic field, $B_0$, set the upstream plasma magnetization parameter, the only free parameter in this study, which is defined as
\begin{equation}
\sigma_0=\frac{B^2_0}{4\pi\Gamma_0 n_0 m_{\rm e}c^2},
\end{equation}
where $m_{\rm e}$ is the electron mass. In this work, we performed runs with $\sigma_0=3$, $10$, $30$, and $100$. Taking the polar profile of $B_{\phi}$ into account and fixing $\chi=\pi/4$ gives an average magnetization, $\sigma$, of the order $\sigma\approx 0.15 \sigma_0$. Therefore, we effectively explored $\sigma=0.45$, $1.5$, $4.5,$ and $15$, meaning from mildly to highly magnetized nebulae. The 3D MHD model favors magnetization of order unity and perhaps higher \citep{2013MNRAS.431L..48P, 2014MNRAS.438..278P}. The physical size of the box is identical in all runs in terms of electron plasma skin depth, $d^0_{\rm e}=(\Gamma_0 m_{\rm e} c^2/8\pi n_0 e^2)^{1/2}$. The spatial resolution is  about five cells per electron skin depth in all directions, therefore fitting about $L_{\phi}=\pi r_{\rm min}/2=320 d^0_{\rm e}$ in the simulation box along the $\phi$ direction.

\section{Results}\label{sect_results}

\subsection{Dynamics and evolution of the shock}

\begin{figure}
\centering
\includegraphics[width=\hsize]{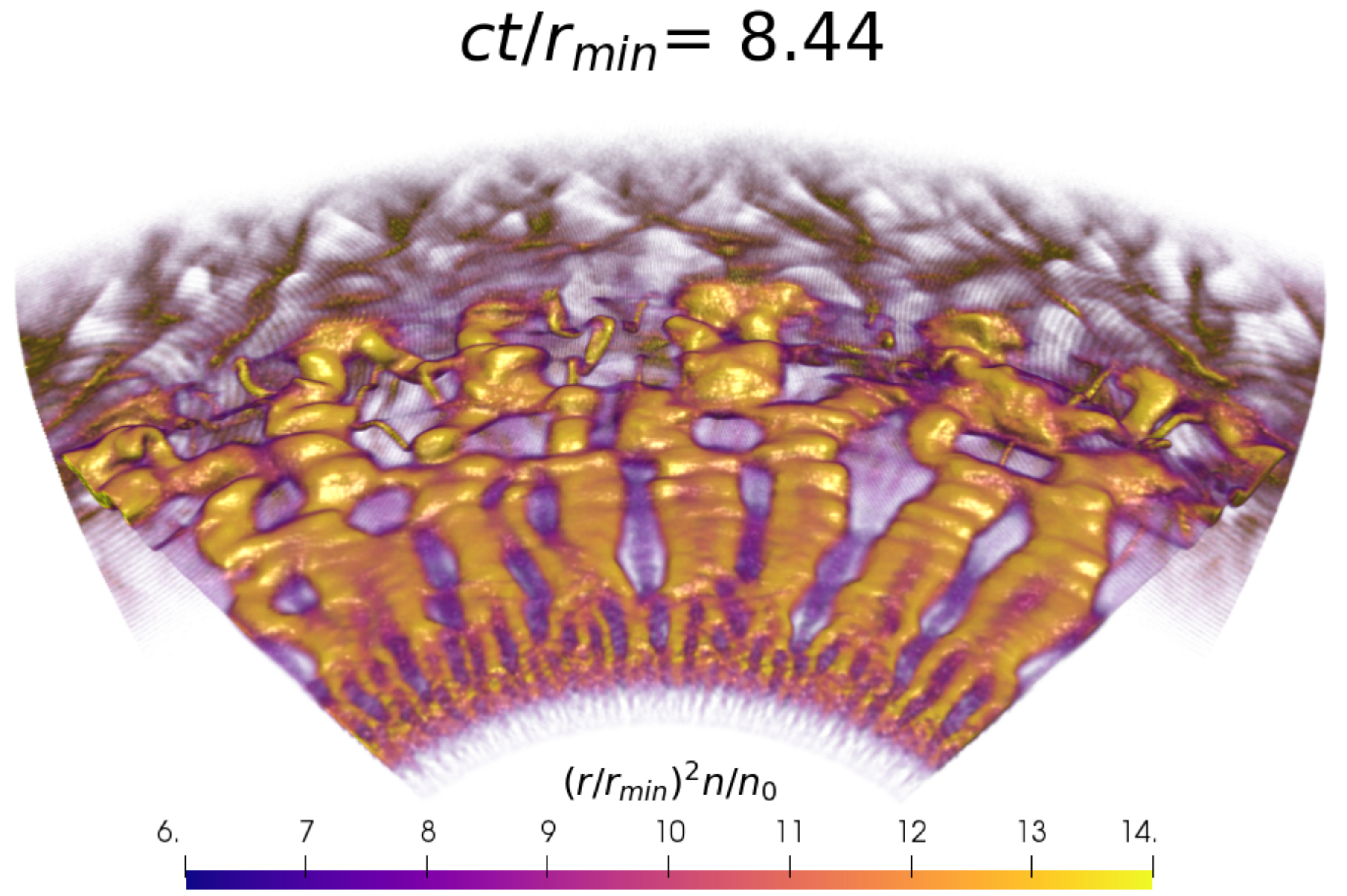}
\caption{Three-dimensional rendering of the final state of the plasma density $(r/r_{\rm min})^2 n/n_0$ for the $\sigma=4.5$ run. For the sake of clarity, axes are not shown; see Fig.~\ref{fig2} for 2D slices with axes and length scales.}
\label{fig1}
\end{figure}

\begin{figure*}
\centering
\includegraphics[width=8.cm]{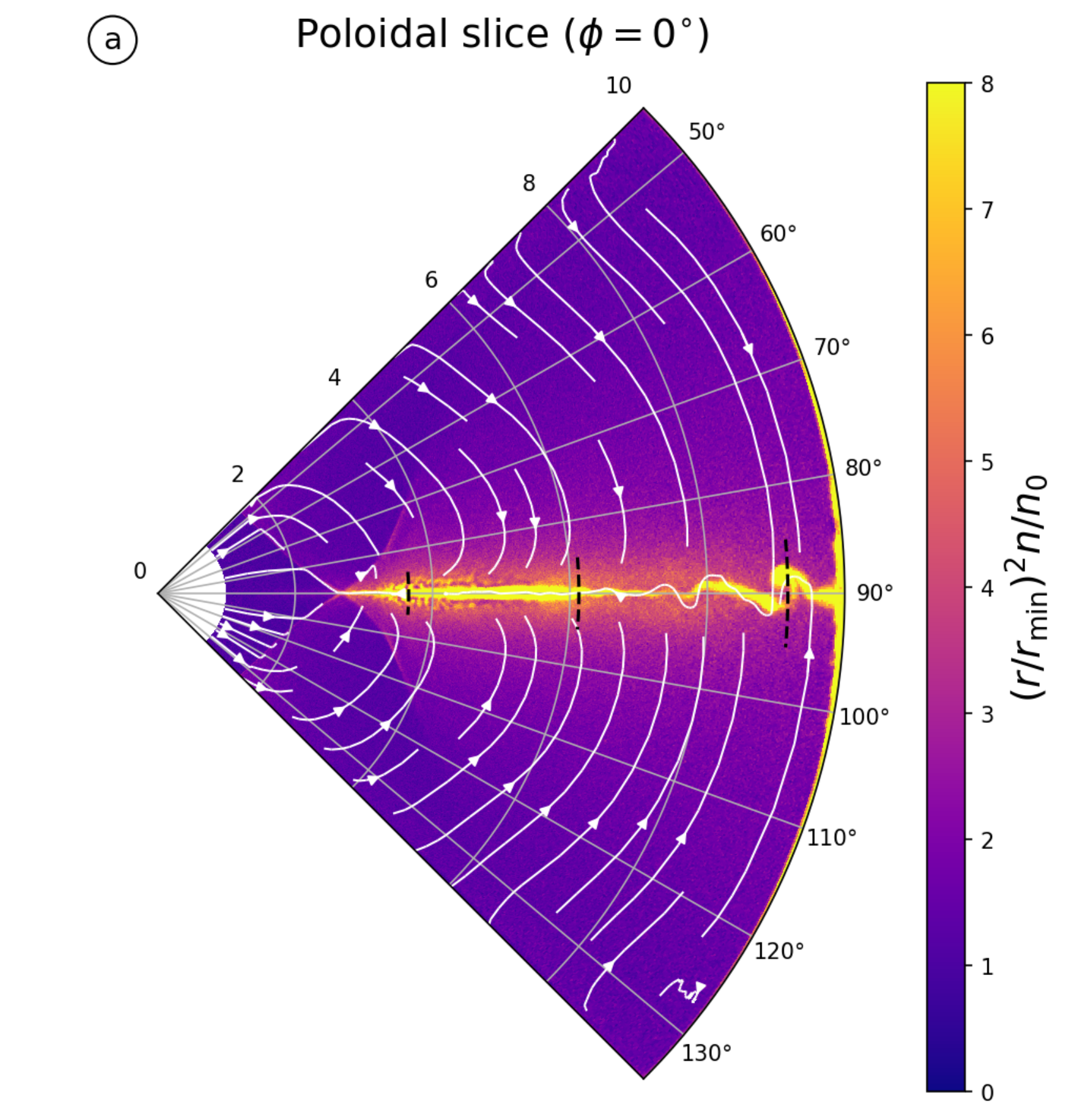}
\includegraphics[width=9.8cm]{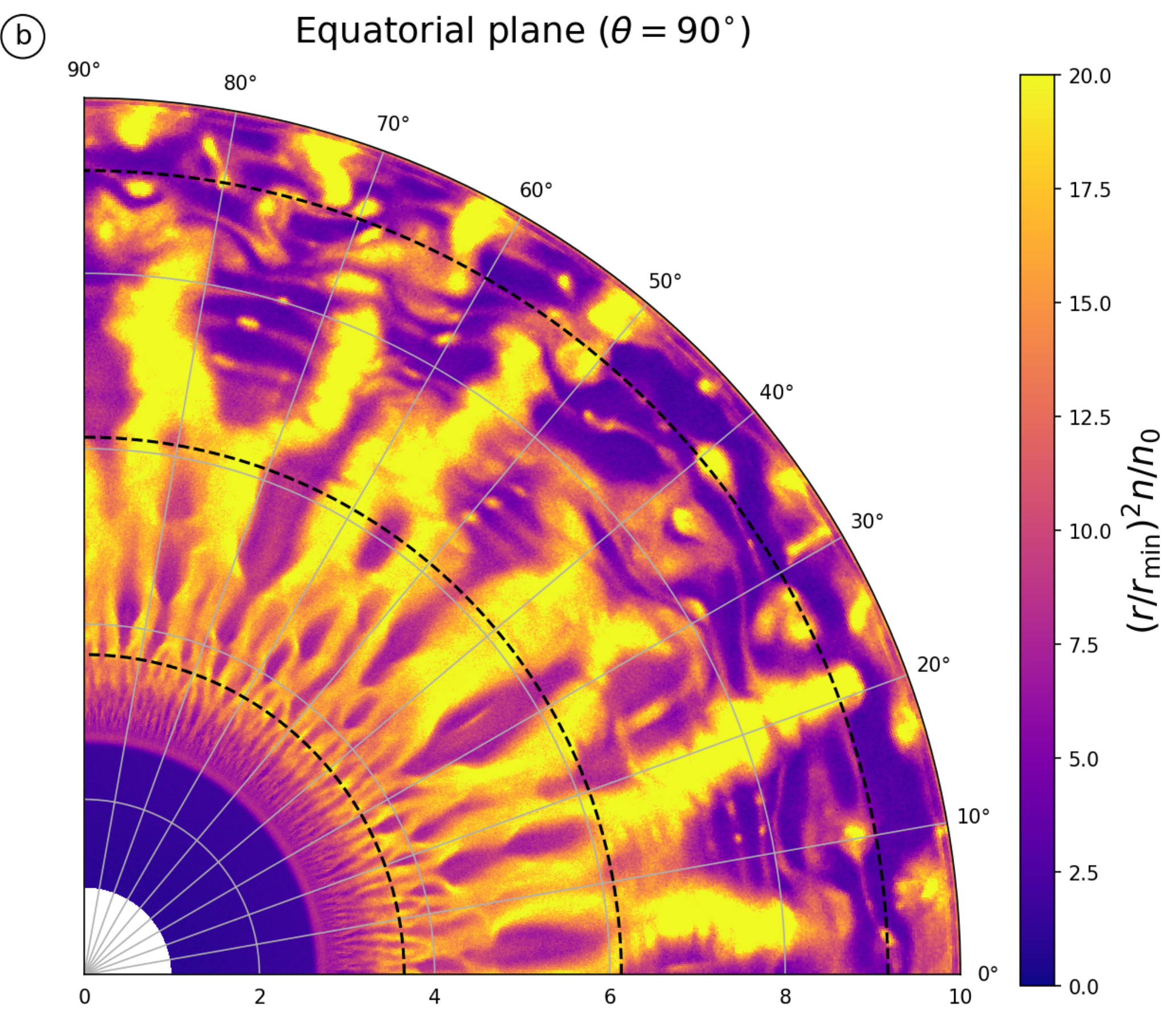}
\includegraphics[width=18cm]{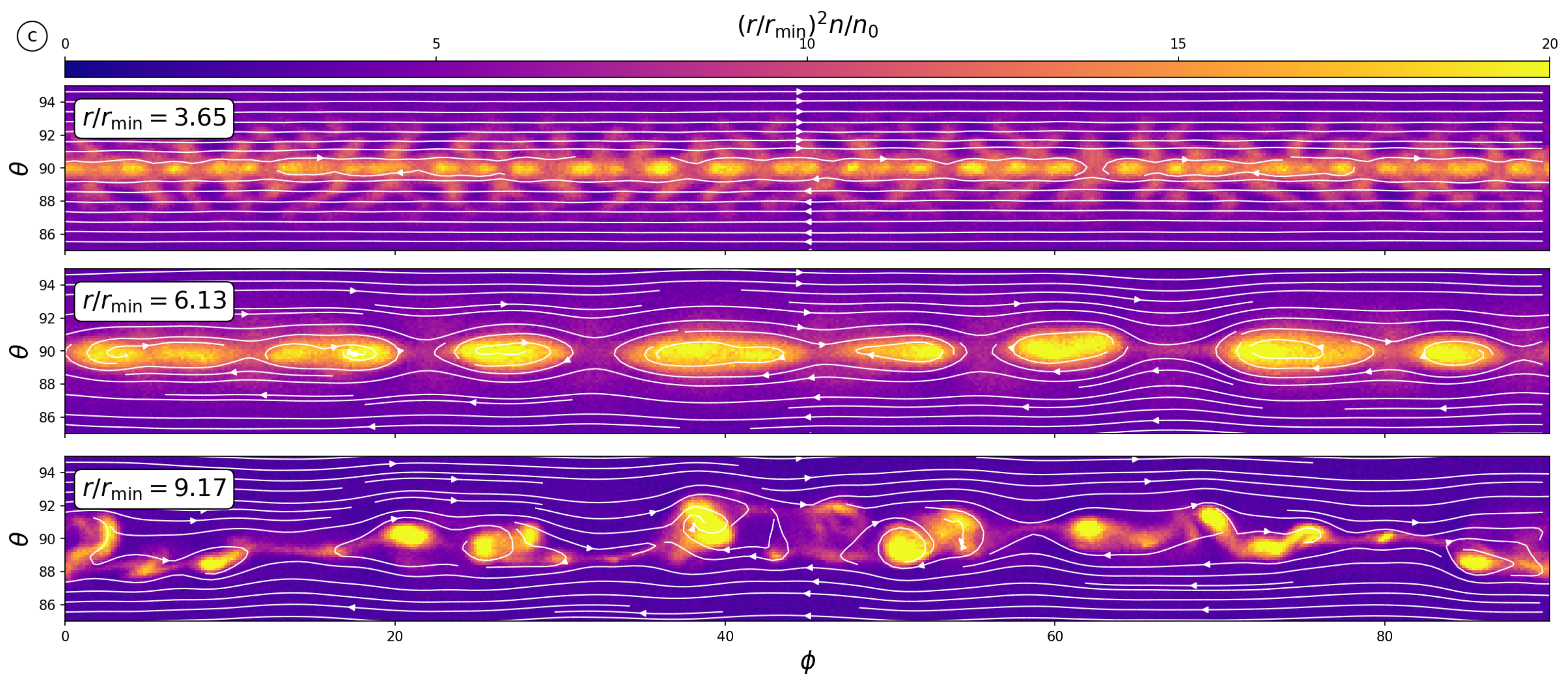}
\caption{Two-dimensional slices through the 3D plasma density map shown in Fig.~\ref{fig1} ($\sigma=4.5$, $t=8.44 r_{\rm min}/c$). Panel {\bf a}: Poloidal cut at $\phi=0^{\circ}$. White solid lines with arrows represent the plasma bulk velocity streamlines. Panel {\bf b}: Equatorial cut ($\theta=\pi/2$). Panel {\bf c}: Cuts through the $\theta\phi$ plane at $r/r_{\rm min}=3.65$, $6.13$, and $9.17$, zoomed-in on the current layer. These radii are indicated by dashed black lines in panels a and b. In panel c, white contours with arrows show the magnetic field lines.}
\label{fig2}
\end{figure*}

The shock forms immediately after the reflection of the beam on the outer boundary. The shock front recedes faster inward as one moves away from the equatorial plane, where the flow is more magnetized and therefore where the shock is weaker, leading to a significant departure from the spherical symmetry of the shock front. The downstream flow is decelerated to mildly relativistic speeds. The most striking feature is the development of a large-scale ring-like current sheet in the equatorial plane due to the sharp magnetic discontinuity forming behind the shock. The downstream flow collapses into the midplane and conducts the electric current via a thin, skin-depth-scale layer. This layer then fragments into multiple interconnected flux ropes along the azimuthal direction, as shown in the 3D rendering of the final plasma density state for $\sigma=4.5$ (Fig.~\ref{fig1}). The current layer is continuously fed with fresh plasma thanks to a large-scale circulation pattern that takes place in the downstream medium and brings high-latitude magnetized plasma down to the midplane (see Fig.~\ref{fig2}, panel a). This mechanism even leads to a reversal of the radial flow velocity in the vicinity of the layer, which in turn excites Kelvin-Helmholtz modes, as reported in \citet{2020A&A...642A.123C} and as conjectured by \citet{1999ApJ...512..755B}. Due to the limited dynamical range probed in these 3D runs, only the early stages of this instability can be captured in the outermost regions of the box ($r/r_{\rm min}\gtrsim 8$; see panel a and the bottom of panel c in Fig.~\ref{fig2}). At later stages, the Kelvin-Helmholtz instability may ultimately dominate and lead to the disruption and transformation of this large-scale current sheet into a turbulent flow \citep{2020A&A...642A.123C}, hence limiting the radial extent of the layer and flux ropes. The shock front cavity reported in \citet{2020A&A...642A.123C} is not observed in these simulations, most likely due to the small system size and integration time.

This network of interconnected flux ropes, reminiscent of 3D plane parallel relativistic reconnection studies (e.g., \citealt{2013ApJ...774...41K, 2014ApJ...782..104C, 2017ApJ...843L..27W, 2021arXiv210602790W}), is clearly visible in the equatorial slice of the plasma density shown in Fig.~\ref{fig2}, panel b. This figure is remarkable in the sense that it captures in a single snapshot the time evolution and the expected merging tree pattern of flux ropes forming in the layer (see for instance the merging tree in the space-time diagram presented in \citealt{2015ApJ...815..101N}). In the $\theta\phi$ plane, the evolution of the layer closely resembles 2D reconnection (panel~c in Fig.~\ref{fig2}). Near the shock front, multiple identical islands form, and this phenomenon corresponds to the linear phase of the tearing instability (at $r/r_{\rm min}\sim 3$ in Fig.~\ref{fig2}, top of panel c). Islands are pushed toward one another by the reconnection outflows and merge, leaving behind a chain of bigger plasmoids separated by secondary layers. These shorter layers become themselves unstable to the secondary tearing mode, and more, but smaller, plasmoids are created ($r/r_{\rm min}\gtrsim 6$; Fig.~\ref{fig1} and panel b of Fig.~\ref{fig2}), which eventually merge with the big islands. The overall dynamics of the shock is robust against magnetization.\ The main differences are the plasma density contrast between the islands, the secondary layers, and the ambient downstream flow that increases with increasing magnetization, meaning that islands are much more visible as structures from the background flow at high magnetization. The strength of the velocity shear across the layer also increases with increasing magnetization, such that the Kelvin-Helmholtz instability is most severe for $\sigma=4.5$ and $15$.

\subsection{Hierarchical merging and formation of giant plasmoids}

\begin{figure}
\centering
\includegraphics[width=\hsize]{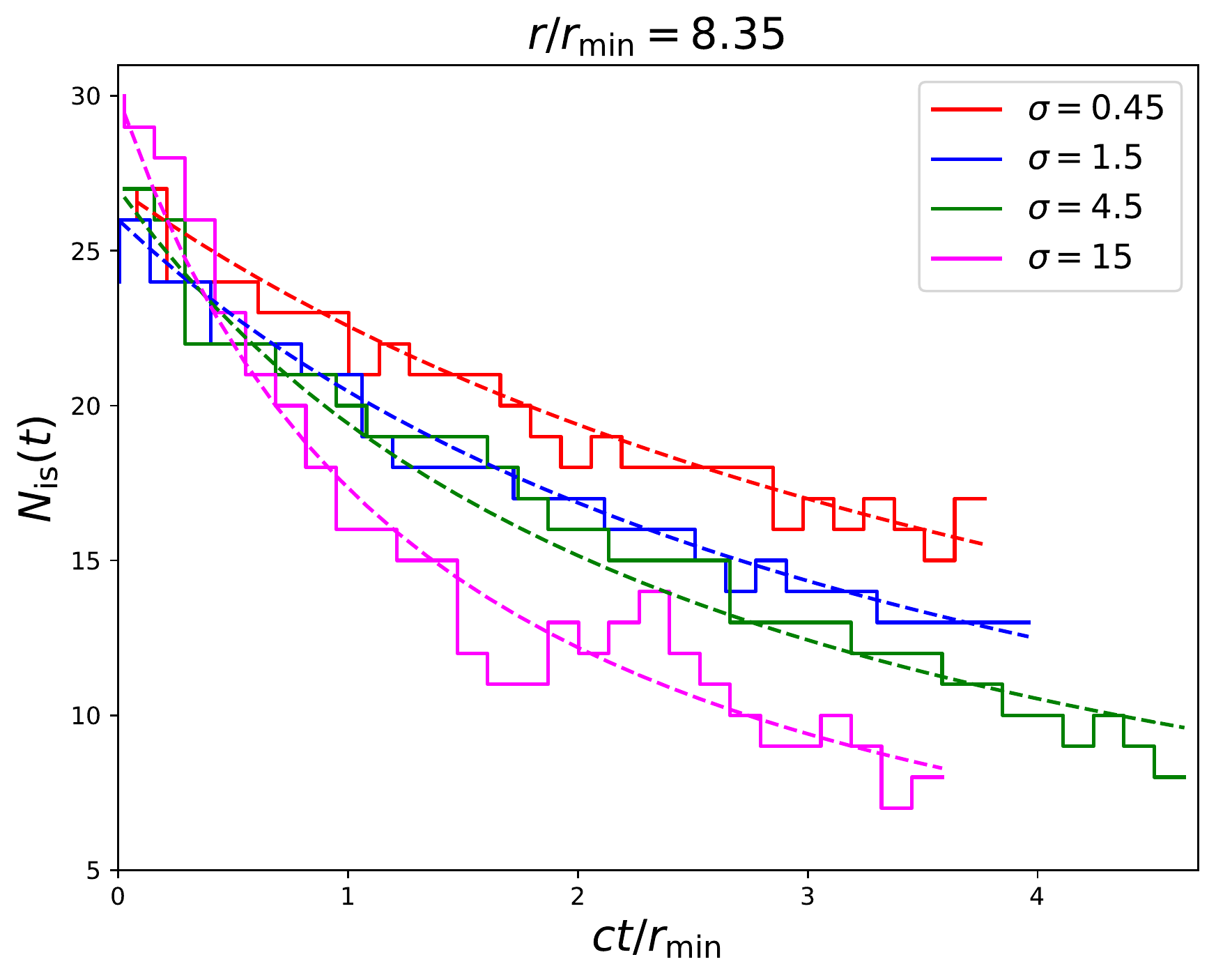}
\caption{Time evolution of the number of plasmoids, $N_{\rm is}$, measured at a fixed radius, $r=8.35 r_{\rm min}$, for each value of $\sigma$ (histograms). The dashed lines represent the best analytical fits to the hierarchical merging model \citep{2019PhRvR...1a2004Z}.}
\label{fig3}
\end{figure}

The merging process inevitably leads to a decrease in the number of plasmoids and to an increase in their sizes (e.g., \citealt{2016MNRAS.462...48S}). To assess whether major plasmoids can grow to macroscopic sizes at the physical scales of pulsar wind nebulae, and therefore examine whether they are viable candidates for explaining the knots, we need to characterize this evolution accurately in the simulations. Figure~\ref{fig3} shows the number of plasmoids as a function of time and at a fixed radius ($r/r_{\rm min}=8.35$). The center of a plasmoid is effectively determined as a local maximum in the magnetic flux function defined in the $\theta\phi$ plane (local minima being the centers of secondary current layers, i.e., X points). In all runs, the initial number of plasmoids is about $30$. Assuming that the layer thickness, $\delta$, is on the order of the electron skin-depth scale, $d_{\rm e}$, as it naturally occurs in a collisionless system, the fastest growing tearing mode has a wavelength of $\lambda_{\rm T}=2\pi\sqrt{3}d_{\rm e}$ \citep{1979SvA....23..460Z, 2007ApJ...670..702Z, 2014ApJ...782..104C}. The ratio of the box size to the tearing wavelength predicts that $L_{\phi}/\lambda_{\rm T}\approx 30$ identical islands should form in the linear phase, in agreement with our results. This number should also be independent of $\sigma$, as reported here, since the box size is a fixed number of electron skin depths. The number of islands then decreases with time and reaches about a dozen toward the end of the simulation. The final number decreases with increasing magnetization from about 16 islands for $\sigma=0.45$ down to about 8 islands for $\sigma=15$. 

In the Crab Nebula, the scale separation between the skin depth and the shock size is much larger than what is achievable with current numerical resources. Assuming a mono-energetic wind of pairs with a Lorentz factor of $\Gamma\sim 10^6$, the number density of particles at the shock is of order $n_{\rm Crab}\sim \kappa L_0/4\pi R^2_{\rm TS}\Gamma m_{\rm e} c^3\approx 5\times 10^{-8}\rm{cm^{-3}}$, where $L_0\approx 5\times 10^{38}\rm{erg/s}$ is the Crab pulsar spin-down power, $R_{\rm TS}\sim 0.1$~parsec is the radius of the termination shock (i.e., the inner ring), and $\kappa=3$ is the shock compression ratio assuming a strong shock. This gives a skin-depth scale of order $d_{\rm Crab}\sim 2\times10^{12}$~cm. Hence, 
\begin{equation}
N^{\rm Crab}_{\rm is,0}\sim \frac{2\pi R_{\rm TS}}{\lambda_{\rm T}}=\frac{R_{\rm TS}}{\sqrt{3}d_{\rm Crab}}\sim 10^5~\rm{islands}
\end{equation}
should be created initially in the Crab Nebula, that is to say, it is about a thousand times larger than what is achieved by 3D PIC simulations ($\sim 4\times 30=120$~islands, if considering a full $2\pi$ domain size). We now address the crucial question of how many islands will remain at the end of the merging process.

To answer this question and fill the gap between our simulation results and the Crab Nebula scale, we made use of the hierarchical model proposed by \citet{2019PhRvR...1a2004Z, 2020JPlPh..86d5301Z, 2021arXiv210413757Z}, which we reproduce here for completeness. This model assumes that the merging mechanism happens in discrete steps, during which identical, equally spaced islands merge by pairs. It also neglects the generation of secondary plasmoids as reconnection proceeds. If the initial number of plasmoids is $N_0$, then their number at step $n$ is
\begin{equation}
N_n=2^{-n}N_0.
\end{equation}
The duration of each merger is controlled by the reconnection timescale given by
\begin{equation}
\tau_n=d_{n}/\beta_{\rm rec}c
\end{equation}
at step $n$, where
\begin{equation}
d_{n}=2^n d_0
\end{equation}
is the half separation between the plasmoids and $\beta_{\rm rec}$ is the dimensionless reconnection rate (i.e., the velocity at which reconnection operates, normalized by the speed of light). The initial half distance between islands is $d_0=L_{\phi}/2N_0$, so the fiducial reconnection timescale is
\begin{equation}
\tau_0=\frac{d_0}{\beta_{\rm rec}c}.
\end{equation}
The total time to reach the nth generation is then
\begin{equation}
t_n=\sum_{k=0}^{n-1}\tau_k=\tau_0\sum_{k=0}^{n-1}2^k=\tau_0\left(2^n-1\right).
\end{equation}
Putting everything together and dropping the subscript $n$ in the continuous limit of multiple successive merging events, the hierarchical model predicts that the number of islands should decay as
\begin{equation}
N_{\rm is}(t)=\frac{N_0}{1+t/\tau_0}.
\label{eq_islands}
\end{equation}
Figure~\ref{fig3} shows that the hierarchical model describes the overall evolution of islands in the simulation well. Leaving the reconnection rate and the initial number of islands as free parameters, we obtain the best-fit solutions, shown in Fig.~\ref{fig3} by dashed lines. The inferred reconnection rate measured at this particular radius ranges from $\beta_{\rm rec}\approx 0.05$ for $\sigma=0.45$ up to $\beta_{\rm rec}\approx 0.15$ for $\sigma=15$. These values fall well within the usual rates reported in the literature (e.g., \citealt{2018MNRAS.473.4840W}).

\begin{figure}
\centering
\includegraphics[width=\hsize]{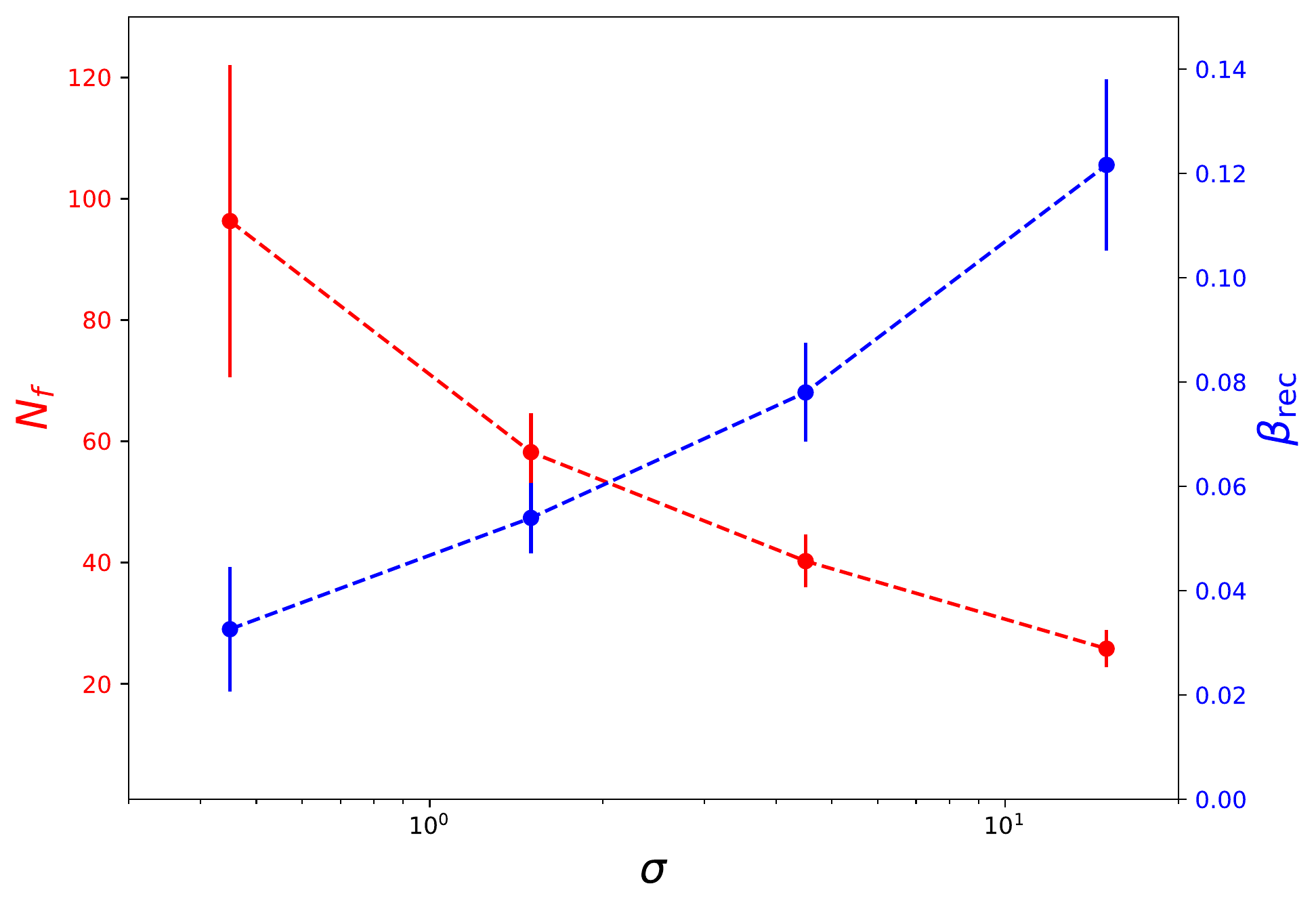}
\caption{Dimensionless reconnection rate, $\beta_{\rm rec}$, as a function of $\sigma$ (blue dots). This rate is measured at multiple locations in the simulation domain, and the error bars show the standard deviation from the mean value. The final number of giant plasmoids is inferred from the hierarchical merging model at the equatorial plane of pulsar wind nebulae, assuming an escape time of $t_{\rm esc}=R_{\rm TS}/c$ (red dots).}
\label{fig4}
\end{figure}

To verify that we are indeed measuring the reconnection rate with this model, we performed a more direct measurement using two diagnostics. The first method consists in measuring the asymptotic inflow velocity toward the major X point, $V_{\rm in}=\beta_{\rm rec}c$, at multiple radii. Once the major X point is identified (using the global minimum of the magnetic flux function), we fit the plasma bulk velocity profile across the layer as $V_{\theta}=V_{\rm in}\tanh\left((\theta-\pi/2)/\delta_{\theta}\right)$, where $\delta_{\theta}\approx\delta/r$ is the angular layer thickness. The second method consists in measuring the ratio of the reconnection electric field inside the layer, here the radial component $E_{\rm rec}$, with the upstream reconnecting magnetic field, here the toroidal component assuming the same profile as for the velocity, $B_{\phi}=B_{\rm up}\tanh\left((\theta-\pi/2)/\delta_{\theta}\right)$, so that $\beta_{\rm rec}=E_{\rm rec}/B_{\rm up}$. Both methods give similar results, but the rate measured with the fields is cleaner (less sensitive to the particle noise) and is reported in Fig.~\ref{fig4}. It ranges from about $\beta_{\rm rec}=0.03$ for $\sigma=0.45$ up to $\beta_{\rm rec}=0.12$ for $\sigma=15$, in good agreement with the rates inferred from the hierarchical model. The increase in the rate with increasing magnetization was also reported in \citet{2018MNRAS.473.4840W}. Beyond $\sigma\gtrsim 10$, the rate should reach a saturation, $\beta^{\infty}_{\rm rec}\gtrsim 0.1$ (see Fig.~9 in \citealt{2018MNRAS.473.4840W}). We note that if the rate is defined as the reconnection speed normalized by the Alfv\'en velocity, $V_{\rm A}=\sqrt{\sigma/(1+\sigma)}$, it would be nearly independent of $\sigma$.

The reconnection rate having been characterized, the last parameter to fix before deducing the final number of giant plasmoids is the total duration of the reconnection process. If the lifetime of the sheet were unlimited, reconnection would saturate and all of the plasmoids would ultimately end up in a single giant island, as seen for instance in simulations with periodic boundary conditions. In pulsar wind nebulae, the lifetime of the layer will be limited by the large-scale dynamics of the post-shock flow. A naive estimate can be given by the advection time of the flow, moving at a speed of order $0.5c$ in the Crab Nebula \citep{2002ApJ...577L..49H}. As mentioned earlier, another possible mechanism is the Kelvin-Helmholtz instability that is due to the strong velocity shear forming across the layer, which could disrupt the large-scale coherent structure of the sheet, converting it into a turbulent flow \citep{2020A&A...642A.123C}. The timescale is on the order of $\sim 0.1 R_{\rm TS}/\Delta V\lesssim R_{\rm TS}/c$ \citep{1982JGR....87.7431M}. Another viable source of disruption is the drift-kink instability \citep{1998ApJ...493..291B, 2011ApJ...728...90M, 2014MNRAS.438..278P}. In this case, the growing time of this instability is also on the order of the light-crossing time of the shock size. All in all, it is an interplay of at least these three effects that will limit the duration of the merging process to about one light-crossing time of the shock size. Thus, defining the escape time as $t_{\rm esc}=R_{\rm TS}/c$, and using the hierarchical model (Eq.~\ref{eq_islands}), the final number of islands is
\begin{equation}
N^{\rm f}_{\rm is}=\frac{N_0}{1+t_{\rm esc}/\tau_{0}}=\frac{N_0}{1+\beta_{\rm rec}R_{\rm TS}/d_0}.
\end{equation}
Given that $\beta_{\rm rec} R_{\rm TS}\gg d_{\rm 0}$, and noticing that $N_0 d_0=L_{\phi}/2=\pi R_{\rm TS}$, this yields
\begin{equation}
N^{\rm f}_{\rm is}=\pi\beta^{-1}_{\rm rec}\approx 30 \beta^{-1}_{0.1,\rm rec}~\rm{islands},
\label{eq_nf}
\end{equation}
where $\beta_{0.1,\rm rec}=\beta_{\rm rec}/0.1$, meaning that the final number of giant plasmoids is solely governed by the reconnection rate, and thus it is remarkably independent of the system size, but also of the initial number of plasmoids (as long as $\beta_{\rm rec} R_{\rm TS}\gg d_{\rm 0}$). Figure~\ref{fig4} reports the corresponding number of islands as a function of $\sigma$ derived from Eq.~(\ref{eq_nf}). The reconnection rate is reported in the same figure. Although the simulations presented here cannot capture the full-scale separation, this important conclusion implies that they are still relevant to capturing the high-end part of the distribution and therefore predicting the correct number of giant plasmoids.

\begin{figure}
\centering
\includegraphics[width=\hsize]{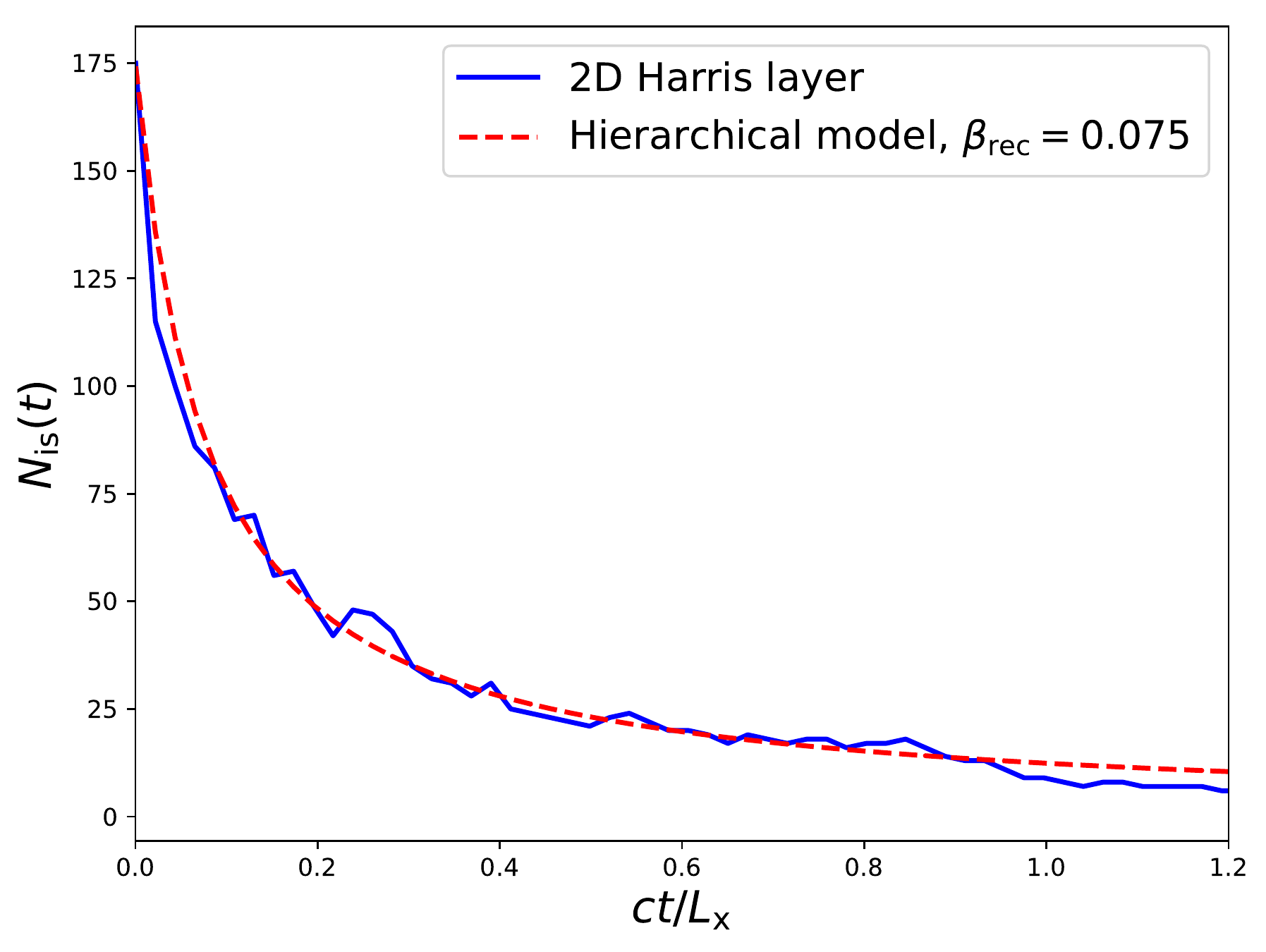}
\caption{Time evolution of the number of plasmoids measured in a 2D Cartesian PIC simulation initialized with a Harris layer of total length $L_x$ (solid blue line), and comparison with the hierarchical model with a best-fit reconnection rate $\beta_{\rm rec}=0.075$ (dashed red line). Small secondary plasmoids forming in the late stages of reconnection between giant plasmoids are not considered in this measurement.}
\label{fig5}
\end{figure}

To further strengthen this point, we performed a standard 2D Cartesian PIC simulation of a Harris reconnection current layer of total length $L_x$ for $\sigma=5$ and with no guide field (i.e., the magnetic component perpendicular to the reconnection plane). Thanks to the larger-scale separation that is achievable in 2D, the initial number of islands is $N_0\approx 175$. This number decays with time in good agreement with the hierarchical model (Fig.~\ref{fig5}). The inferred rate is $\beta_{\rm rec}\approx 0.075$, which is compatible with the rate measured in the $\sigma=4.5$ 3D run. The number of plasmoids after one light-crossing time of the layer, $L_x/c$, is $N^{\rm f}_{\rm is}\approx 20$, which is comparable to the 3D shock simulation despite their initial number differing by a factor of $\approx 6$. 

\subsection{The effect of expansion}

In the above discussion, the effect of expansion has not been explicitly taken into account. Yet, measurements of proper motions in the Crab Nebula show that the downstream flow is expanding at mildly relativistic speeds, as also expected from the shock jump conditions \citep{1984ApJ...283..694K}. Simulations presented here do not capture this effect well, despite the spherical geometry, because of the static reflecting outer boundary condition chosen for numerical convenience. Flux ropes are produced nearly at rest downstream of the shock as the latter recedes inward, and therefore they do not experience significant expansion over time.

To assess the effect of expansion, we considered a collection of $N_{\rm is}$ equally spaced islands co-moving in the radial direction at a constant velocity, $v_{\rm exp}$. Each closest pair of islands is separated by an angle $\alpha=2\pi/N_{\rm is}$. In the co-moving frame, islands only have a motion along the azimuthal direction (i.e., perpendicular to the flow); this is set by the reconnection speed, which in the lab frame is $\beta_{\rm rec}c/\Gamma_{\rm exp}$, where $\Gamma_{\rm exp}$ is the Lorentz factor of the expanding flow. Islands will be able to merge if the speed at which they are pushed away from each other by expansion (i.e., $\alpha v_{\rm exp}$) is smaller than the reconnection speed,
\begin{equation}
\frac{2\pi}{N_{\rm is}}v_{\rm exp}\lesssim\frac{\beta_{\rm rec}c}{\Gamma_{\rm exp}}.
\end{equation}
Assuming a mildly relativistic expanding flow, $\Gamma_{\rm exp}\approx 1$, which is applicable to the nebula, the number of islands should not be smaller than
\begin{equation}
N^{\rm f}_{\rm is}\gtrsim\frac{2\pi v_{\rm exp}}{\beta_{\rm rec}c}.
\end{equation}
Taking $v_{\rm exp}/c=0.5$ for the Crab gives $N^{\rm f}_{\rm is}\gtrsim\pi\beta^{-1}_{\rm rec}$ islands, which is the same result we get from Eq.~(\ref{eq_nf}). In other words, even if the current layer had an infinite lifetime, it would never be able to reach a fully saturated state where only a single giant plasmoid would remain, in contrast to a static and finite system. Therefore, the spherical expansion of the downstream flow truncates the tail of the distribution predicted by the hierarchical model. This causality argument gives an additional and robust constraint in favor of the expected number of giant flux ropes formulated in Eq.~(\ref{eq_nf}), which was obtained by limiting the lifetime of the hierarchical merging process to the light-crossing time of the shock.

\subsection{Synchrotron hotspots}

\begin{figure}
\centering
\includegraphics[width=\hsize]{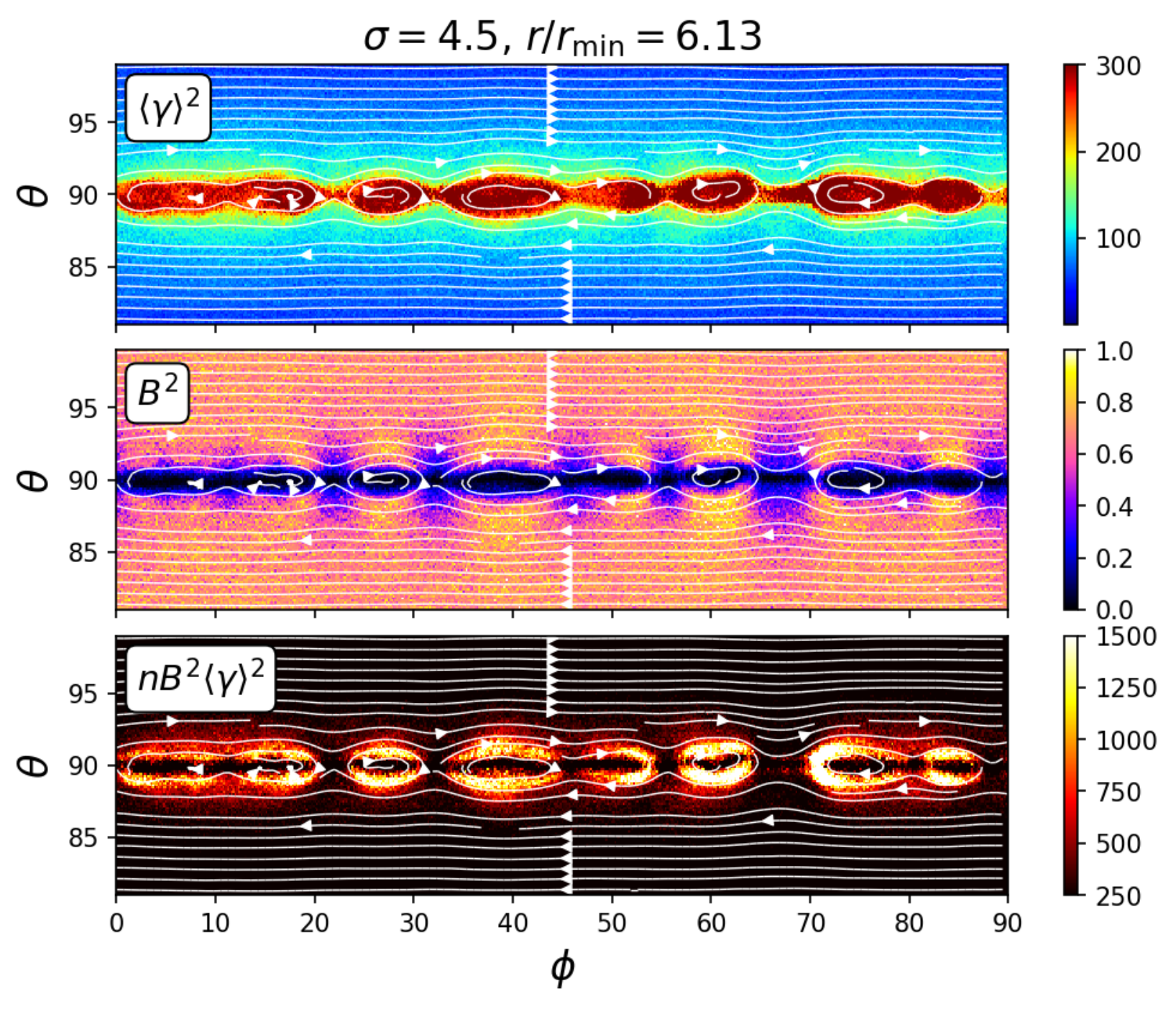}
\caption{Zoomed-in view on the current sheet sliced at $r/r_{\rm min}=6.13$ for the $\sigma=4.5$ run. Top panel: Square of the mean particle Lorentz factor, $\langle\gamma\rangle^2$. Middle panel: Square of the total magnetic field strength, $B^2$. Bottom panel: Proxy for the total synchrotron power, $n B^2\langle\gamma\rangle^2$. White contours show the magnetic field lines projected on this plane.}
\label{fig6}
\end{figure}

Giant magnetic islands are filled with all of the particles energized by reconnection up to a particle Lorentz factor $\gamma\sim \Gamma_0 \sigma$, and hence they are intense sources of nonthermal synchrotron radiation and appear as bright knots or, more precisely, as filaments downstream of the shock front along the equatorial plane (Fig.~\ref{fig6}). Although synchrotron emission is highly polarized and the magnetic field in islands well ordered, the polarization signal averaged over the surface of the island for a loop-like magnetic structure should be canceled out, leaving a mostly unpolarized signal. This would provide a natural explanation as to why knots are not visible in polarized optical light \citep{2008ARA&A..46..127H}, in contrast to the inner knot. The observed flux variability of the knots can also be accounted for by the well-known dynamical nature of the flux rope network. With a few dozen knots in the inner ring in the Crab, this model thus favors a magnetized $\sigma\gtrsim 5$ pulsar wind nebula \citep{2010MNRAS.405.1809L, 2014MNRAS.438..278P, 2020A&A...642A.123C}. This conclusion is independently supported by the hard spectral index measured along the inner ring in X-rays (amongst the hardest values found in the Crab Nebula; \citealt{2004ApJ...609..186M}). With a reported X-ray spectral index of $\alpha\sim 0.8$ in the ring ($F\propto \nu^{-\alpha}$, where $\nu$ is the photon frequency and $F$ the flux of photons per second), this yields an injected particle distribution of $dN/d\gamma\propto \gamma^{-1.6}$. Such a hard particle distribution can be achieved by reconnection for $\sigma\gtrsim 5$ \citep{2014PhRvL.113o5005G, 2016ApJ...816L...8W}.

\section{Conclusion}

Pulsar wind nebulae decidedly offer a formidable and clean environment for studying relativistic magnetic reconnection in nature. In this work we have shown that a large-scale ring-like current sheet forms and reconnects in the equatorial plane downstream of pulsar wind termination shocks. Reconnection operates in a nearly ideal textbook-like configuration: a thin, flat layer with periodic boundary conditions (along the azimuthal direction) and no guide field that is continuously reforming behind the shock. It is fed by the high-latitude magnetized downstream flow. Reconnection proceeds in the relativistic plasmoid-dominated regime, leading to an efficient fragmentation of the current layer, dissipation of the toroidal magnetic field, and nonthermal particle acceleration. A large number of small, nearly identical plasmoids are first created in the immediate vicinity of the shock front. They merge in a hierarchical manner, leaving behind a few macroscopic structures that contain all the energetic particles processed by reconnection. Assuming that reconnection is unperturbed for at most a light-crossing time of the shock radius -- after which Kelvin-Helmholtz and kink instabilities will mix and disrupt the layer, leading to a turbulent state -- the final number of giant plasmoids is solely determined by the inverse of the dimensionless reconnection rate, $N^{\rm f}_{\rm is}=\pi\beta^{-1}_{\rm rec}$. A similar conclusion can be drawn from a causality argument: plasmoids can merge until the expansion of the downstream flow freezes their evolution. The minimum final number of large flux ropes depends on the ratio of the expansion to the reconnection speeds.

This remarkable result predicts that only magnetized nebulae such as the Crab ($\sigma\gtrsim 5$) should present discernible large islands (a few dozen at most); otherwise, they would appear as a continuous ring rather than a collection of well-defined discrete entities. We argue that the presence of these giant plasmoids (i.e., flux ropes) is a robust feature of pulsar wind termination shocks and provides a natural explanation as to the mysterious knots observed along the inner ring in the Crab Nebula (which, in fact, should appear more like filaments rather than circular structures). Hence, counting the knots gives a direct measure of the reconnection rate and average plasma magnetization injected by the wind into the nebula. In this interpretation, the presence of knots represents a real smoking gun of the magnetic reconnection process, which could well be involved in the Crab gamma-ray flare activity in the reconnection scenario \citep{2011ApJ...737L..40U, 2013ApJ...770..147C}. Another important implication of this work is that the shock front should not coincide with the inner ring; instead, it should be located closer inward and should be under-luminous in comparison with the giant plasmoids forming further downstream.

To make more progress and capture these knots in a fully global setup along with the other features of the nebula (such as the jet, the wisps, and the inner knot, which are missing here), high-resolution 3D resistive MHD simulations with test particles could be used to provide an adequate answer before this becomes feasible with full PIC simulations. Last, we would like to point out that the same argument as the one developed in this work can be applied to other astrophysical environments where a large-scale current sheet forms, such as the striped wind current sheet in pulsars (e.g., \citealt{2020A&A...642A.204C}) and the ergospheric current sheet in black hole magnetospheres \citep{2019PhRvL.122c5101P, 2021A&A...650A.163C}, with perhaps significant observational consequences to come.

\begin{acknowledgements}
The authors would like to thank Benjamin Crinquand, Lorenzo Sironi, and the referee Andrei Bykov for valuable comments regarding this work, and Gilles Henri for drawing our attention to the effect of expansion. This project has received funding from the European Research Council (ERC) under the European Union’s Horizon 2020 research and innovation programme (grant agreement No 863412). Computing resources were provided by TGCC and CINES under the allocation A0090407669 made by GENCI.
\end{acknowledgements}

\bibliographystyle{aa}
\bibliography{knots_pwn}

\end{document}